\documentclass[npg]{copernicus}

\def\bq{\begin{equation}}
\def\eq{\end{equation}}
\begin{document}

\title{Non-diffusive, non-local  transport in fluids and plasmas}
\author{D. del-Castillo-Negrete}
\affil{Oak Ridge National Laboratory}


\runningtitle{Non-diffusive, non-local  transport in fluids and plasmas}

\runningauthor{D. del-Castillo-Negrete}

\correspondence{D. del-Castillo-Negrete\\ delcastillod@ornl.gov}
\received{}
\pubdiscuss{} 
\revised{}
\accepted{}
\published{}

\firstpage{1}

\maketitle
\begin{abstract}
A review of non-diffusive transport in fluids and plasmas is presented. In the fluid context, non-diffusive chaotic transport by Rossby waves in zonal flows is studied following a Lagrangian approach. In the plasma physics context the problem of interest is test particle transport in pressure-gradient-driven plasma turbulence. In both systems the probability density function (PDF) of particle displacements is strongly non-Gaussian and the statistical moments exhibit super-diffusive anomalous scaling.  
Fractional diffusion models are proposed and tested  in the quantitative description of the  non-diffusive Lagrangian statistics of the fluid and plasma problems. Also, fractional diffusion operators are used to construct non-local transport models exhibiting 
up-hill transport, multivalued flux-gradient relations, fast pulse propagation phenomena, and ``tunneling" of perturbations across transport barriers. 
\end{abstract}

\introduction


The widely used advection-diffusion equation rests on the validity of the Fourier-Fick's prescription which in the case of transport of a single scalar, $T$, in a one-dimensional domain states that, $q=-\chi \partial_x T+  V T$, where $q$ is the flux, $\chi$ is the diffusivity, and $V$ the advection velocity. From the statistical mechanics point of view,  this model assumes an underlying Markovian, Gaussian, uncorrelated stochastic process. However, despite the relative success of the diffusion model, there are cases in which this model fails to describe transport, and an alternative description must be used.  The goal of this paper is to review some  recent results on non-diffusive transport of particular interest to fluids and plasmas.  We focus on non-diffusive Lagrangian particle transport and non-local transport of passive scalar fields. 

In  the paradigmatic case of the Brownian random walk, the Gaussian statistics of the individual particle displacements, and the lack of correlations and memory effects (Markovian assumption), lead to a Gaussian PDF of the net particle displacement, and to the linear  in time scaling for the mean, $M \sim t$, and the variance, $\sigma^2 \sim t$. Based on these scaling, the   transport coefficients are defined as $V=\lim_{t \rightarrow \infty} M(t)/t$ and $\chi=\lim_{t \rightarrow \infty} \sigma^2(t)/t$. 
 The signatures of non-diffusive behavior in Lagrangian particle transport include non-Gaussian PDFs of particle displacements and anomalous scaling of the moments of the form  $M \sim t^\xi$ and  $\sigma^2 \sim t^\gamma$, with $\xi \neq 1$ and/or $\gamma \neq 1$.  When $\gamma>1$ ($\gamma<1$)  transport is super-diffusive (sub-diffusive),
 see for example \cite{bouchard}. In either case, the diffusion model cannot be applied because the effective diffusivity $\chi$ is either $\infty$ or zero.

The study of non-diffusive Lagrangian particle transport presented here focuses on two systems of interest to geophysical fluid dynamics and plasma physics. In the geophysical context we consider transport in quasigeostrophic zonal flows. 
Quasigeostrophic flows are $2$-dimensional, rapidly rotating flows in which there is a gradient in the Coriolis force. These flows are relevant in the study of mesoscale dynamics in the atmosphere and the oceans, see for example  \cite{pedlosky}.  Zonal shear flows occur naturally in nature; two well-known examples are the Gulf Stream and the polar night jet above Antarctica. Barotropic perturbations of these flows give rise to low frequency instabilities known as Rossby waves that have a key influence on the dynamics and transport. Following \cite{del_castillo_1993,del_castillo_1998} we study chaotic transport by Rossby waves in zonal shear flows.  In the plasma physics context we consider non-diffusive transport in pressure-gradient-driven plasma turbulence. This system is of relevance to the understanding of magnetically  confined fusion plasmas.
In this case, the Lagrangian particle dynamics corresponds to the motion of test particles in the presence of an external fixed magnetic field and a fluctuating turbulent electrostatic potential. 
In the fluid and the plasma physics problems, we present numerical evidence of non-diffusive transport. In particular, in both cases, the PDFs of particle displacements are strongly non-Gaussian and the variance exhibits anomalous scaling of the super-diffusive type.

As mentioned before, when the statistical moments exhibit anomalous scaling, the advection-diffusion model can not be applied and alternative models must be used. In this paper we review the use of fractional derivatives to construct such alternative models. Fractional derivatives are integro-differential operators that generalize the concept of derivatives of order $n$, to fractional orders \cite{samko_1993,podlu_1999}. Although the origins of fractional calculus go back to the origins of regular calculus, the use of fractional derivatives in the mathematical modeling of transport  is relative recent.  We  present a brief review of this formalism in connection with the continuous time random walk (CTRW) model. The CTRW  generalizes the  Brownian random walk by incorporating non-Gaussian jump PDFs and non-Markovian waiting time PDFs 
\cite{montroll_weiss,montroll_shlesinger,metzler}.  Following this, we construct effective macroscopic fractional diffusion models of the PDFs of particle displacements 
\cite{del_castillo_2004,del_castillo_2005}. A comparison is presented between  the analytical solutions of the fractional models and the numerical results obtained from the Lagrangian statistics for the fluid and plasma problems mentioned above.  

The use of fractional derivatives in transport modeling is close related to the problem of nonlocal transport. By nonlocal we mean that the flux of the transported scalar at a point depends on the gradient of the scalar throughout the entire domain. The generic mathematical structure of the nonlocal flux is $q = -\chi \int {\cal K}(x-y) \partial_y T dy $, where the function 
${\cal K}$ measures the degree of nonlocality.  The ``width" of this function depends on the strength of the non-locality, and in the limit when  ${\cal K}$ is a Dirac delta  function, the flux reduces to the local Fourier-Fick's prescription.  
Motivated by the successful use of fractional derivatives to model non-diffusive Lagrangian transport, we discuss the use
of these operators to construct non-local model of passive scalar transport. 
Following \cite{del_castillo_2006,del_castillo_2008} we present numerical results illustrating important non-local transport phenomenology including: up-hill transport, multivalued flux-gradient relations, fast pulse propagation phenomena, and ``tunneling" of perturbations across transport barriers.  

The rest of this paper is organized as follows. Section~II discusses non-diffusive chaotic transport by Rossby waves in zonal flows. Non-diffusive turbulence transport in plasmas is studied in Sec.~III. Section~IV presents a brief review of fractional diffusion. The applications of fractional diffusion to model the PDFs of particle displacements in the  Rossby waves and the plasma problems are discussed in Sec.~V. Non-local transport is studied in Sec.~VI, and Sec.~VII presents the conclusions.

\section{Non-diffusive chaotic transport by Rossby waves in zonal flows}

In this section we study non-diffusive chaotic transport by Rossby waves in zonal shear flows. Since the flow is $2$-dimensional and incompressible, the flow  velocity can be written as  
${\bf v}=(-\partial_y \Psi, \partial_x \Psi)$ where $\Psi(x,y,t)$ is the streamfunction. In this case, the Lagrangian trajectories of individual tracers, $d{\bf r}/dt={\bf v}$, are obtained from the solution of the Hamiltonian system,
\bq
\label{tracers}
\frac{d x}{dt}=-\frac{\partial \Psi}{\partial y} \, \qquad 
\frac{d y}{dt}=\frac{\partial \Psi}{\partial x} \, . 
\eq
where $\Psi$ plays the role of the Hamiltonian and the ${\bf r}=(x,y)$ spatial coordinates play the role of canonically conjugate phase space coordinates. Hamiltonian systems of the form in Eq.~(\ref{tracers}) are always integrable when $\Psi$ does not depend on time. However, when $\Psi$ depends explicitly on time, the system can be non-integrable and individual trajectories can be chaotic, see for example  \cite{tabor}. The main goal of the study of chaotic transport is to understand the global transport properties of tracers in this case, see for example  \cite{otino}.  Problems of particular interest to geophysical flows include the study of the formation and destruction of transport barriers \cite{del_castillo_1993}, and the study of the Lagrangian statistics
\cite{del_castillo_1998}. Here we focus on the second problem. 

To construct a model for the streamfunction $\Psi(x,y,t)$ we have to consider the dynamics of the system. 
In the case of quasigeostrophic flows, 
$\Psi(x,y,t)$ is obtained from the potential vorticity conservation law  
\bq
\label{pv}
\frac{\partial q}{\partial t} +\left( {\bf v} \cdot \nabla \right) q=0 \, ,
\eq
where according to the  $\beta$-plane approximation, $q=\nabla^2 \Psi + \beta y$. 
We have adopted a right-handed Cartesian coordinate system with $z$ pointing in the direction of the rotation of the system and $y$ in the direction of the Coriolis force gradient. That is, $y$ points in the ``northward" direction and $x$ is a periodic coordinate in the ``eastward" direction.  

To simplify the solution of the non-linear Eq.~(\ref{pv})  we assume a streamfunction of the form
\bq
\label{psi_model}
\Psi=\Psi_0(x,y)+\Psi_1(x,y,t) \, ,
\eq
where $\Psi_0$, is the superposition of a zonal shear flow with dependence $u_0(y)= \tanh y$, and 
a  regular neutral  mode  in 
 its co-moving reference frame,
\bq
\Psi_0= -\ln \left( \cosh y \right) + \epsilon_1 \phi_1(y) \cos (k_1 x) + c_1 y \, .
\eq
The function $\Psi_1$ is a time dependent perturbation of the form
\bq
\Psi_1=\epsilon_2 \phi_1(y) \cos (k_1 x-\omega t) \, ,
\eq
where $\epsilon_1$ and $\epsilon_2$ are free parameters determining the amplitude of the linear Rossby waves, and $\omega$ is the frequency of the perturbation.
The eigenfunction $\phi_1$,
\bq
\phi_1=\left[ 1 + \tanh y \right]^{(1-c_1)/2}\, \left[ 1 - \tanh y \right]^{(1+c_1)/2} \, , 
\eq
is obtained from the solution of the linear eigenvalue problem of the quasigeostrophic equation and 
$(k_1,c_1)$ are obtained from the corresponding dispersion relation for neutral (zero growth rate) modes
\cite{del_castillo_1998}.

When $\epsilon_2=0$ the streamfunction is time independent and the solution of Eq.~(\ref{tracers}) can be reduce to a quadrature. 
In this case the Lagrangian dynamics is integrable and the  orbits of the tracers can be classified in two types: (i) trapped orbits that encircle  the vortices and (ii) untrapped orbits that move freely in the East-West direction following the  zonal shear flow. These two types of orbits are separated by the separatrix that joints the hyperbolic stagnation points of the flow. 
When $\epsilon_2 \neq 0$, the system ceases to be integrable. In particular, as shown in Fig.~1, the perturbation breaks the separatrix and creates a stochastic layer where tracers alternate chaotically between following the zonal flow and being trapped inside the vortices. 

To characterize transport in the chaotic regime  we follow a statistical approach. The most basic quantity is the probability density function (PDF) of particle displacements. Transport in the ``north-south" direction  is trivial since particle orbits in the $y$-direction are bounded by the zonal flows. Therefore, we focus on transport in the ``east-west" direction, i.e. along the zonal flow. Given an ensemble of initial conditions, $\{ (x_i, y_i)\}$ with $i=1,\,2,\, \ldots N_p$ we compute the PDF of particle displacements, $P(\delta x, t)$ where $\delta x_i(t)=x_i(t)-x_i(0)$.  By definition, at $t=0$ the PDF is a Dirac delta function, $P(\delta x, t=0)=\delta(\delta x)$. As $t$ increases, the PDF widens and might drift to one side or the other.  Note that, although $\delta x$ is a periodic function in the annular domain shown in Fig.1, to compute the statistics we treat $\delta x$ as variable defined on the $(-\infty, \infty)$ domain. 

\begin{figure}[t]
\label{testu}
\vspace*{2mm}
\begin{center}
\includegraphics[width=6.0cm]{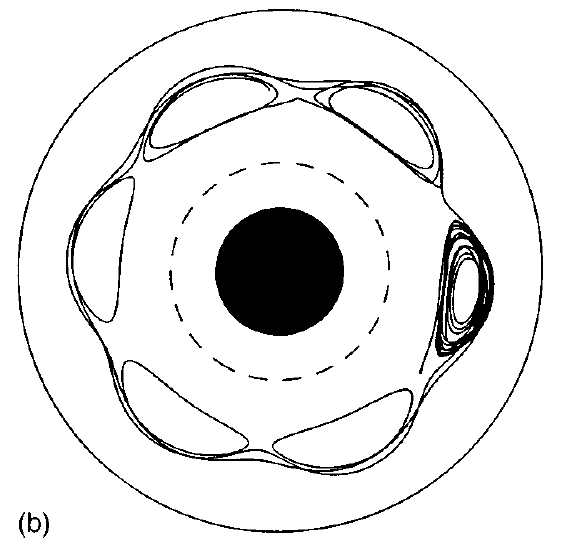}
\end{center}
\caption{Chaotic transport by Rossby waves in the quasigeostrophic zonal flow in Eqs.~(\ref{tracers}) and (\ref{psi_model}).  In the presence of two or more Rossby waves, the trajectories of passive tracers are typically chaotic. In particular, as shown in the figure, tracers alternate in a seemly random way between being trapped in the vortices and moving freely along the ``east-west", $x$ angular direction, following the shear flow flanking the vortices.}
\end{figure}

\begin{figure}[t]
\label{fig_fluid_pdf}
\vspace*{2mm}
\begin{center}
\includegraphics[width=6.0cm]{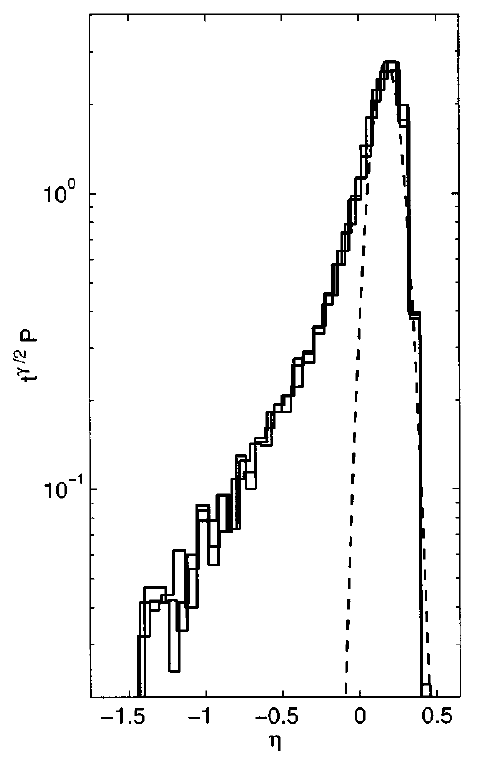}
\end{center}
\caption{Rescaled probability distribution function (PDF), $t^{\gamma/2} P$, of passive tracers displacements, $\delta x(t)=x(t)-x(t=0)$, as function of the similarity variable, $\eta=\left( \delta x- \langle \delta x\rangle \right)/ t^{\gamma/2}$ with $\gamma=1.9$.  
The dynamics corresponds to  the quasigeostrophic model in Eqs.~(\ref{tracers}) and (\ref{psi_model}). 
The plot shows the PDF at $t=800$, $900$ and $1000$. Consistent with the self-similar scaling in Eq.~(\ref{scaling}), the PDFs at successive times collapse. The anomalously large displacements induced by the zonal flow (see Fig.~1) result in the strong departure of the $\eta<0$ tail from the Gaussian fit (dashed line). The value $\gamma>1$ indicates super-diffusive transport. }
\end{figure}

To study the self-similar evolution of the PDF we introduce  the scaling variable 
\bq
\label{similarity_var}
\eta= \langle \delta x - \langle \delta x \rangle \rangle\, t^{-\gamma/2} \, . 
\eq
Figure~2 shows the rescaled PDF, $t^{\gamma/2} P$, as function of $\eta$. The observation that  the  rescaled PDFs collapse for successive times leads support to the assumption that, at large times, $P$ converges to a self-similar distribution of the form
\bq
\label{scaling}
P^*(x,t)=t^{-\gamma/2} f(\eta) \, ,
\eq
where $f$ is a scaling function, and $\gamma$ is the scaling exponent. 
The scaling in Eq.~(\ref{scaling})  implies the following scaling of the moments
\bq
\label{moments}
\langle X^n \rangle \sim t^{n \gamma/2} , ,
\eq
where $X= \delta x - \langle \delta x \rangle$.  Equation~(\ref{scaling}) also implies
\bq
P^*(X,t)=\lambda^{\gamma/2} P^* \left( \lambda^{\gamma/2} X, \lambda t \right)\, ,
\eq
where $\lambda$ is a free parameter. According to this relation, up to an scale factor, the limit distribution, $P^*$, is invariant under the space-time renormalization operation $(X, t) \rightarrow \left( \lambda^{\gamma/2} X, \lambda t \right)$. That is, the PDF at a later time can be obtained from a rescaling of the PDF at an early time. 

In the diffusive case, $P^*$ is a Gaussian, $\gamma=1$, and Eq.~(\ref{scaling}) corresponds to the 
similarity solution of the advection-diffusion equation. However, in the numerical results shown in Fig.~2, transport is non-diffusive because $\gamma \neq 1$ and the scaling function is not a Gaussian. In particular, the tails of the PDFs exhibit a decay significantly slower than Gaussian and a strong asymmetry. 
 Because, $\gamma > 1$, it is concluded that ``East-West", azimuthal chaotic transport by Rossby waves is zonal flows is super-diffusive. For further details on the statistics and a  dynamical explanation of the dependence of the  asymmetry of the PDF on the perturbation frequency $\omega$ see  Ref.~\cite{del_castillo_1998}. This reference also discusses the comparison of the model presented here with  experimental results on transport in rapidly rotating fluids 
\cite{solomon_1993}.

It is interesting to mention that there is a very close analogy between the dynamics of Rossby waves in rapidly rotating neutral fluids in the quasigeostrophic approximation and drift-waves in magnetized plasmas, see for example
\cite{petvia,horton,horton_book}. 
In this analogy, the role of the rapid rotation is played by the strong magnetic field, the 
 fluid streamfunction corresponds to the electrostatic potential,  
the fluid vorticity to the plasma density, and the gradient in the Coriolis force corresponds to the plasma density background gradient.  Based on this analogy, as discussed in \cite{del_castillo_2000}, the results presented here are directly applicable to the study of non-diffusive chaotic transport by drift waves in magnetized plasmas.

\section{Non-diffusive turbulent transport in plasmas}

\begin{figure}[t]
\label{fig_plasma_eddies}
\vspace*{2mm}
\begin{center}
\includegraphics[width=6.0cm]{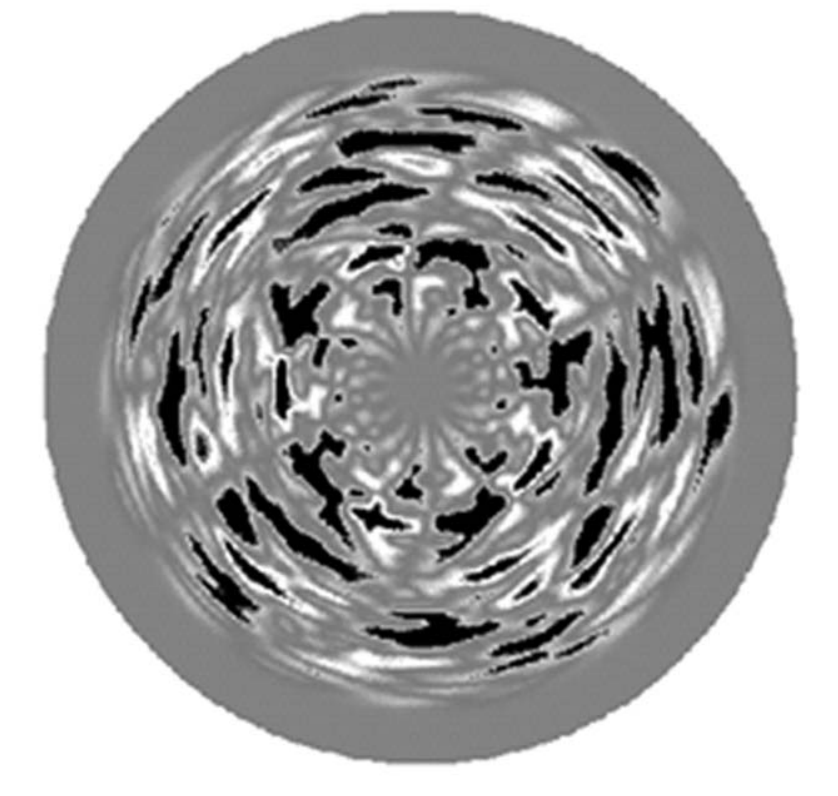}
\end{center}
\caption{Fluctuating electrostatic potential ${\tilde \Phi}$ at a fixed time obtained from the numerical integration of the plasma turbulence model in Eqs.~(\ref{model_1})-(\ref{model_2_2}). The dark (light) coherent patches denote rotating (counter rotating) ${\bf E} \times {\bf B}$ eddies.  The trapping effects of these eddies along with intermittent large radial displacements caused by avalanche-like plasma relaxation events, give rise to non-diffusive transport and to the non-Gaussian PDF in  Fig.~4 \cite{del_castillo_2004,del_castillo_2005}.}
\end{figure}

\begin{figure}[t]
\label{fig_PDF_plasma}
\vspace*{2mm}
\begin{center}
\includegraphics[width=8.0cm]{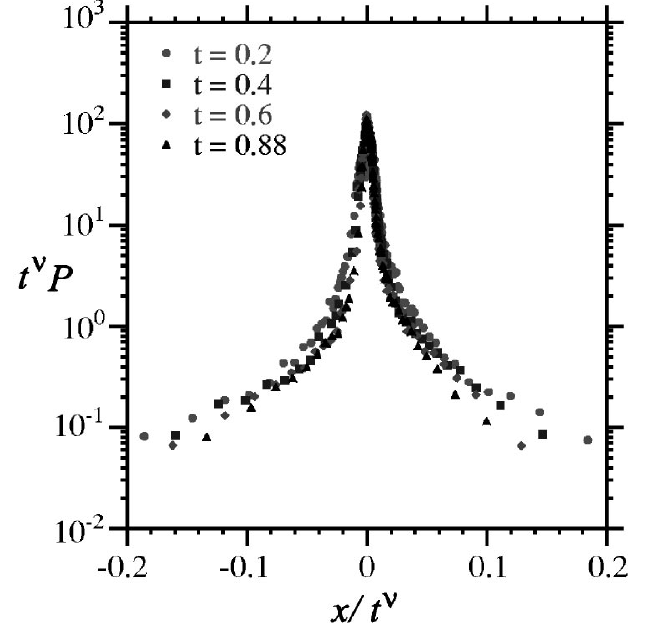}
\end{center}
\caption{Rescaled probability distribution functions (PDFs), $t^{\nu} P$, of passive tracers radial displacements, $x(t)$, as function of the similarity variable, $x/ t^{\nu}$ with $\nu=2/3$.  
The dynamics correspond to  the pressure-gradient-driven plasma turbulence model in Eqs.~(\ref{model_1})-(\ref{model_2_2}). 
The plot shows the PDF at $t=0.2$, $0.4$, $0.6$ and $0.88$. Like in the fluid dynamic case in Fig.~2,  the collapse of the PDFs at successive times indicates a self-similar scaling of the form in Eq.~(\ref{scaling}). In this case, the non-diffusive transport manifest in the slowly decaying non-Gaussian tails of the PDF. The value $\nu>1/2$ indicates super-diffusive transport \cite{del_castillo_2004,del_castillo_2005}. 
}
\end{figure}

In the example discussed in the previous section, transport resulted from chaotic advection. That is, from the chaotic dynamics of the deterministic equations describing the particle orbits. In particular, the streamfunction $\Psi$ is a deterministic differentiable function. In the case of turbulent transport the situation is different since the flow velocity advecting the tracers is a nondeterministic, random function. 
Nevertheless, turbulent systems can also exhibit non-diffusive transport of passive tracers. 
In this section we present an example in the context of plasma physics.  

As in the previous section, we follow a Lagrangian approach and consider the statistics of a large ensemble of tracer particles. In the plasma, the particle motion responds to the combined effect of a turbulent electric field, 
$\tilde{\bf E}=- \nabla \tilde{\Phi}$, and a fixed external magnetic field, ${\bf B}_0$. 
The equation of motion of the tracers are obtained from Newton's law with the Lorentz force. However, 
in the guiding center approximation, see for example \cite{nicholson},  the equations can be simplified as the first order system
\bq
\label{guiding_center}
\frac{d {\bf r}}{dt}=\frac{1}{B^2_0} \nabla \tilde{\Phi} \times {\bf B}_0 \, ,
\eq
where ${\bf r}=(x,y)$ denotes the position of the particle in the $2$-dimensional plane perpendicular to the magnetic field. 
This system has also a Hamiltonian structure with the potential, $ \tilde{\Phi}$, playing the role of Hamiltonian.

The fluctuating plasma electrostatic potential, $ \tilde{\Phi}$, is  obtained from the solution of the turbulence model. Here,
following \cite{del_castillo_2004,del_castillo_2005}, we consider pressure-gradient-driven turbulence in cylindrical geometry.
The underlying instability of this type of turbulence is the resistive interchange mode, driven by the pressure gradient. This instability is the analogue of the Rayleigh-Taylor instability responsible for the gravity-driven overturning  of high density fluid laying above a low density fluid. In magnetically confined plasmas, the role of gravity is played by the magnetic field lines curvature.
The turbulence model \cite{carreras_1987} is  based on an electrostatic
approximation of the reduced  resistive magneto hydrodynamic equations, 
\bq
\label{model_1}
\frac{d}{dt}\nabla_\perp^2\,
\tilde{\Phi}
 =
-\frac{1}{\eta m_i n_0 R_0}\nabla^2_\parallel \tilde \Phi+\frac{B_0}{m_i n_0}\frac{1}{r_c}
\frac{1}{r}\frac{\partial \tilde p}{\partial \theta}+ \mu \nabla^4_\perp \tilde \Phi \, ,
\eq
\bq
\label{model_2}
\frac{d}{dt}\tilde p=
\frac{\partial \left< p\right>}{\partial r} \frac{1}{r}\frac{\partial \tilde \Phi}{\partial \theta}
+\chi_\perp \nabla^2_\perp \tilde p+ \chi_\parallel \nabla_\parallel^2  \tilde p \, ,
\eq
where $\tilde \Phi$ is the electrostatic potential, 
$\tilde{p}$ the pressure, and $d/dt=\partial_\tau + \tilde{\bf{V}}\cdot \nabla$.
The instability drive is the flux-surface averaged pressure gradient, $\partial \langle
p\rangle/\partial r$, determined according to  
\bq
\label{model_2_2}
\frac{\partial \left<p\right>}{\partial \tau}+ \frac{1}{r} \frac{\partial }{\partial r}r \left< \tilde V_r
\tilde p\right>= S_0 + D \frac{1}{r}\frac{\partial}{\partial r}\left ( r \frac{\partial \left<
p \right>}{\partial r}\right) \, .
\eq
The tildes indicate fluctuating quantities (in space and time), and the angular brackets,
$\left< \, \right>$, denote flux surface averaging over a cylinder at a fixed radius.  
The equilibrium density is $n_0$, the ion mass is $m_i$, the averaged radius of curvature of the
magnetic field lines is $r_c$, and the resistivity is $\eta$.  
The sub indices  ``$\perp$" and ``$\parallel$" denote the 
direction perpendicular and parallel to the magnetic field respectively.
The function $S_0$ represents a source of particles  and heat which we model using a parabolic 
profile, $S_0=\bar{S}_0\left[1-(r/a)^2\right]$.  Figure~3 shows a snapshot in time of the fluctuating electrostatic potential $\tilde \Phi$ obtained form the solution of Eqs.~(\ref{model_1})-(\ref{model_2_2}). 
 
Having computed $ \tilde{\Phi}$, the next step is to integrate Eq.~(\ref{guiding_center}) to obtain the orbits of the tracers. 
The initial condition consists of $25 \times 10^3$ particles with random initial positions in $\theta$ and $z$, and radial position $r=0.5 a$. By definition, at $t=0$, the PDF, $P$, of radial particle displacements, $x=[r(t)-r(0)]/a$, is a Dirac delta function. As time advances the $P(x,t)$, spreads and develop slowly decaying, ``fat" tails. Figure~4 shows the long-time time behavior of the PDF as function of the similarity variable $x/t^\nu$. The strong non-Gaussianity of $P$ is evident. Like in the previous fluid example case, transport is super-diffusive because $\nu >1/2$. 
Evidence of non-diffusive transport has also been observed in other plasma systems including
gyrokinetic simulations of ion temperature gradient (ITG) turbulence \cite{sanchez_et_al_2008}. 

\section{Fractional diffusion models of  non-diffusive transport}


One of the main goals of transport modeling is to construct effective macroscopic transport equations that reproduce experimentally or numerically observed phenomena. 
For example, in the fluid and the plasma transport problems discussed in the previous two sections, 
the goal is to construct a transport equation that describes the observed spatio-temporal evolution of the PDF, $P$, of particle displacements. 

When transport is diffusive, a simple solution to this problem is provided by  the advection-diffusion equation 
\bq
\label{adv_diff}
\partial_t P + V \partial_x P= \partial_x \left( \chi \partial_x P \right) \, ,
\eq
where the advection velocity and diffusivity are obtained from the asymptotic behavior the statistical moments 
\bq
V = \lim_{t \rightarrow \infty} \frac{\langle x(t) \rangle}{t} \, , \qquad 
\chi = \lim_{t \rightarrow \infty} \frac{\langle x^2(t) \rangle}{2 t} \, ,
\eq
of the particle's displacements, $x$. 
However, this approach fails in the case of non-diffusive transport. In particular, according to the scaling in Eq.~(\ref{moments}) when there is super-diffusion, $\chi \rightarrow  \infty$. Moreover, as it is well-known, the Green's function of Eq.~(\ref{adv_diff}) in an unbounded domain, is a translated Gaussian and this  significantly limits the range of PDFs that this model  can describe. In particular, PDFs with $\gamma \neq 1$ scaling and/or with slowly decaying tails, like those obtained in the examples discussed before 
(Figs.~2 and 4),  cannot be modeled using a simple advection-diffusion equation. 


From the statistical mechanics point of view, the advection-diffusion model assumes an underlying Markovian, Gaussian stochastic process with a drift, i.e. a biased Brownian random walk, see for example \cite{stochastics_book}. However,  the description of transport in the presence of coherent structures requires the use of random walk models that incorporate more 
general stochastic processes. In particular, in the fluid problem discussed in Sec.~2, the trapping effect of the vortices gives rise to non-Markovian effects, and the zonal shear flows give rise to non-Gaussian particle displacements. 
In the plasma physics problem discussed in Sec.~3, the non-Markovian effects are due to the trapping in electrostatic eddied, and the non-Gaussian particle displacements result from avalanche-like radial relaxation events. 

The Continuous Time Random Walk  (CTRW) model
\cite{montroll_weiss,montroll_shlesinger,metzler} 
provides an elegant powerful framework to incorporate these type of effects. 
The CTRW generalizes the Brownian walk in two ways. First,
contrary to the Brownian random walk where particles are assumed to jump at
discrete fixed time intervals, the CTRW model allows the possibility
of incorporating a waiting time probability distribution, $\psi(t)$.  In addition, the CTRW model allows the
possibility of using
non-Gaussian jump distribution functions, $\eta(x)$, with divergent  moments to 
account for long displacements known as L\'{e}vy flights.
Given $\psi$ and $\eta$,
the probability of finding a tracer at position $x$ and time $t$ is
determined by the Montroll-Weiss master equation 
\begin{equation}
\label{eq_3_1}
\partial_t P = \int_0^t dt'\, \phi(t-t')  \int_{-\infty}^\infty dx' \left[
\eta (x-x') P(x',t') - \right.
\eq
$$\left. - \eta (x-x') P(x,t') \right ]  \, ,
$$
The spatial integral on the 
right-hand-side  represents the
gain-loss balance for $P$ at $x$. In particular, the
first term inside the square bracket gives the increase of $P$ due
to particles moving to $x$ while the second term describes the
decrease of $P$ due to particles moving away from $x$. The time integral accounts for memory effects weighted by the function $\phi(t)$. 
In Fourier-Laplace variables,
\bq
\label{ft_lt_defs}
{\cal F} \left[ \eta \right] =\hat{\eta}(k)=\int_{-\infty}^\infty 
e^{ikx} \eta(x) dx \, ,
\eq
\bq
{\cal L} \left[ \phi \right] =\tilde{\phi}(s)=\int_{0}^\infty e^{st} 
\phi(t) dt \, ,
\eq
Eq.~(\ref{eq_3_1}) takes the  form
\begin{equation}
\label{eq_MW_FL}
\hat{\tilde{P}}(k,s)=\frac{1-\tilde{\psi}(s)}{s}\frac{1}{1-\tilde{\psi}(s)\hat{\eta}(k)}
\, .
\end{equation}
where  the relation between the waiting time PDF and the memory 
function is $\tilde{\phi}= s 
\tilde{\psi}/\left(1-\tilde{\psi}
\right)$.

The Montroll-Weiss master Eq.~(\ref{eq_3_1}) can be used directly to model non-diffusive transport, see for example 
\cite{van_milligen_2004,spizzo}. 
However,  this description carries in a sense too much information concerning the details of the underlying stochastic process that might be irrelevant in the long-time, large-scale description of transport. This motivates the 
derivation of  a macroscopic transport equation from Eq.~(\ref{eq_MW_FL}) valid in the time asymptotic ($s \rightarrow 0$) long-wavelength
($k \rightarrow 0$) ``continuum" limit 
\cite{saichev_1997,metzler,scalas_etal_2004}. 
A key aspect of this limit is that only the asymptotic behavior,
i.e., the tails of the $\eta$ and $\psi$ PDFs matter.  This
is a significant advantage over the use of the kinetic master
equation that requires the detailed knowledge of these functions. 

As expected, in the Markovian-Gaussian case
\bq
\label{m_7}
\psi(t)= \mu\,  e^{-\mu\, t}\, , \qquad \eta(x)=\frac{1}{\sqrt{2 \pi} 
\sigma}\,
e^{-x^2/(2
\sigma^2)}\, ,
\eq
where $\langle t \rangle=1/\mu$ is the characteristic waiting time and
$\sigma^2=\langle x^2\rangle$ is the characteristic mean square jump, 
the fluid limit
of the master equation Eq.~(\ref{eq_MW_FL}) leads to the standard 
diffusion equation in (\ref{adv_diff}).
However, the situation is quite different in the case of algebraic 
decaying PDFs of the form
\bq
\label{m_12}
\psi \sim t^{-(\beta+1)}\, , \qquad \eta \sim |x|^{-(\alpha +1)} \, ,
\eq
where for simplicity we have assumed that $\eta$ is symmetric.
In this case, for  $0<\beta<1$,  $\langle t \rangle$ diverges, and there 
is no characteristic
waiting time. Similarity, for $\alpha <2$, $\langle x^2 \rangle$ 
diverges, indicating   a lack of
characteristic transport scale.
The use of this type of algebraic decaying PDFs
is motivated by the  significant probability of very large  trapping 
events and very large spatial displacements, as it is the case in the examples discussed in Secs. 2 and 3.
From the asymptotic behavior in Eq.~(\ref{m_12}) it follows that for small $s$ and $k$, 
\bq
\label{m_13}
\tilde{\psi}(s) \approx  1 - s^\beta + \ldots \, , \qquad 
\hat{\eta}(k)  \approx 1- |k|^\alpha + \ldots \,
\eq
Substituting Eq.~(\ref{m_13}) into Eq.~(\ref{eq_MW_FL}) we get to leading order
\bq
\label{m_14}
s^\beta \, \hat{\tilde{P}}(k,s) - s^{\beta-1} = -\chi \, |k|^\alpha\, 
\hat{\tilde{P}}(k,s) \, .
\eq
To obtain the macroscopic transport equation we 
need to invert the Fourier-Laplace transforms in
Eq.~(\ref{m_14}). This can be formally done by writing
\bq
\label{frac_diff_model}
_0^cD_t^\beta \, P = \chi\,  D_{|x|}^\alpha P\, ,
\eq
where the operators in Eq.~(\ref{frac_diff_model}) are defined according to
\bq
\label{m_15}
{\cal L}\left[ _0^cD_t^\beta \, P \right] =s^\beta \, \tilde{P}(x,s) 
- s^{\beta-1}\, \delta(x)\, \, ,
\eq
\bq
\label{m_15_2}
{\cal F}\left[ D_{|x|}^\alpha  \, P \right] =-|k|^\alpha \, \hat{P}(k,t) \, ,
\eq
for $0<\beta<1$.
Equations~(\ref{m_15}) and (\ref{m_15_2}) are the natural generalizations of the Laplace transform of a time derivative
and the Fourier transform of a spatial derivative. This motivates the formal identification of
the operator
$_0^cD_t^\beta$ as a ``fractional time derivative" for  $0<\beta<1$, and the operator 
$D_{|x|}^\alpha$  as a ``fractional space derivative" for $1<\alpha<2$. As expected, for $\alpha$ or $\beta$ integers, the regular derivatives are recovered. 

The previous discussion assumed a symmetric jump stochastic process, $\eta(x)=\eta(-x)$. 
It can be shown that in the general case the transport equation is
\bq
\label{frac_diff}
_0^cD_t^\beta \, P = \chi \left[ l \, _{-\infty}D_x^{\alpha} + r \, _xD_{\infty}^{\alpha} \right] \, P 
\, ,
\eq
where the operators on the right hand side are the left and right Riemann-Liouville fractional derivatives of order $\alpha$
\cite{samko_1993,podlu_1999}
\bq
\label{eq_3_30}
_{a}D_x^\alpha f = \frac{1}{\Gamma(m-\alpha)}\,
\frac{\partial^m}{\partial x^m}\,
\int_a^x\, \frac{f(y)}{(x-y)^{\alpha+1-m}}\, dy \, ,
\eq
\bq
\label{eq_3_31}
_{x}D_b^\alpha f = \frac{(-1)^m}{\Gamma(m-\alpha)}\,
\frac{\partial^m}{\partial x^m}\,
\int_x^b\, \frac{f(y)}{(y-x)^{\alpha+1-m}}\, dy \, ,
\eq
where $m$ is a positive integer such that $m-1 \leq \alpha < m$. 
In this general formulation, the asymmetry of the underlying stochastic process manifests on the parameters $l$ and $r$,
 \bq
\label{eq_3_29}
l= - \frac{(1-\theta)}{2 \cos(\alpha \pi /2)} \, , \qquad
r= - \frac{(1+\theta)}{2 \cos(\alpha \pi /2)}\, ,
\eq
that  control the relative weight of the left and right fractional derivatives,
where $-1\leq \theta \leq 1$. In the symmetric case, $\theta=0$, 
$D_{|x|}^\alpha = \frac{-1}{2 \cos \left( \pi \alpha/2 \right)}\left[
\,_{-\infty}D_{x}^\alpha+\,_{x}D_{\infty}^\alpha  
\right ]$ which corresponds to the operator defined in Fourier space in Eq.~(\ref{m_15_2}).
In the time domain, the fractional derivative operator in time, $_{0}^{c}D_t^\beta$, introduced in Eq.~(\ref{m_15}) become an integro-differential operator of the form
\begin{equation}
\label{eq_3_39}
_{0}^{c}D_t^\beta P= \frac{1}{\Gamma(1-\beta)} \, \int_0^t
\frac{\partial_{t'} P}{(t-t')^\beta}\,d t' \, ,
\end{equation}
where $0<\beta <1$.
For a derivation of fractional diffusion models that incorporate more general stochastic processes, including the physically important case of truncated L\'evy statistics, see \cite{cartea_del_castillo_2007}.
For a derivation of fractional diffusion models using quasi-linear type renormalization techniques see
\cite{sanchez_et_al_2006}.  

\section {Applications of Fractional diffusion models}

The goal of  this section is to use the fractional diffusion equation to model the non-diffusive transport of tracers discussed in Secs.~2 and 3. In particular, we show that the  numerically obtained PDFs of the particle displacements in Figs. 2 and 4 can be obtained as solutions of effective macroscopic fractional diffusion equations.  

The solution of the initial value problem of Eq.~(\ref{frac_diff}) 
with  $P(x,t=0)=P_0(x)$ is
\bq
\label{gen_sol}
P(x,t) = \int_{-\infty}^\infty P_0(x') G (x-x',t) dx' \, ,
\eq
where the Green's function (propagator)  $G$ is the solution of the 
initial value problem
$G(x,t=0)=\delta(x)$ with $\delta(x)$  the Dirac delta function.
Using Eqs.~(\ref{m_15}) and (\ref{m_15_2}),
the Fourier-Laplace transform of Eq.~(\ref{frac_diff}) leads to the solution
\bq
\label{fl_g}
\hat{\tilde G}= \frac{s^{\beta-1}}{s^\beta -  \Lambda(k)}\, ,
\eq
where
\bq
\label{psi_kk}
\Lambda =\chi \left[ l (-ik)^\alpha + r(ik)^\alpha \right]\, , 
\eq
for $\alpha \neq 1$.
Introducing the Mittag-Leffler function, see for example \cite{podlu_1999},
\bq
\label{ml_fcn}
E_\beta(z)=\sum_{n=0}^{\infty} \frac{z^n}{\Gamma(\beta n+1)}\, , \qquad
{\cal L}\left[  E_\beta(c \,t^\beta) \right]=
\frac{s^{\beta-1}}{s^\beta -c} \, ,
\eq
the inversion of  the Fourier-Laplace transform in Eq.~(\ref{fl_g}) gives
\bq
\label{green_fcn}
G(x,t)=t^{-\beta/\alpha} \, K (\eta )\, , 
\eq
\bq
\label{k_fcn}
K(\eta)= \frac{1}{2 \pi} \int_{-\infty}^{\infty} e^{-i \eta k}\,
E_\beta \left[ \Lambda (k) \right ]
d k\, ,
\eq
where
\bq
\eta=x (\chi^{1/\beta} t )^{-\beta/\alpha}
\eq
is the similarity variable.
Further details of the solution of the initial value problem and 
useful asymptotic and convergent expansions of the
Green's function can be found in 
Refs.~\cite{metzler,saichev_1997,mainardi_etal_2001}.

Of particular interest  is the asymptotic behavior in $x$, 
 for a fixed $t=t_0$,
\bq
\label{x_alpha_scaling}
G (x,t_0)\sim x^{-(1+\alpha)} \, , \qquad x\gg \left( \chi_f^{1/\beta}
t_0\right)^{\beta/\alpha}
\, .
\eq
and the small $t$ and large $t$ scaling at 
fixed $x=x_0$,
\bq
\label{t_beta_scaling}
G(x_0,t)\sim \left\{
\begin{array}{ll}
t^\beta & {\rm for} \qquad   t  \ll \left( \chi_f^{-1}\, x_0^{\alpha}
\right)^{1/\beta}
\\ \\
t^{-\beta} &  {\rm for} \qquad   t  \gg
\left( \chi_f^{-1}\, x_0^{\alpha}
\right)^{1/\beta} \, .
\end{array}
\right.
\eq
From these relations it follows that the order of the fractional derivative in space, $\alpha$,
determines the algebraic asymptotic scaling of the  propagator
in space for a fixed time, and
the order of the fractional derivative in time,  $\beta$, determines the
asymptotic algebraic scaling of the propagator in time for a fixed $x$.
These two properties  provide a  useful guide to construct
fractional models given the spatio-temporal asymptotic scaling
properties of  the PDF.
 Using Eq.~(\ref{green_fcn}), the
moments in the fractional model are given by 
\bq
\label{mo}
\langle x^n \rangle = \int x^n\, P(x,t) \, dx \sim t^{n \beta/\alpha}
\int  \eta^n\,   K(\eta) \, d\eta \, ,
\eq 
that implies the anomalous diffusion scaling
\bq
\label{gamma_ab}
\langle x^2 \rangle \sim t^\gamma \, , \qquad 
\gamma= 2 \beta/\alpha \, .
\eq

According to Fig.~2, the scaling exponent of the PDF of particle displacements in chaotic transport by Rossby waves is 
$\gamma \sim 1.9$.  As expected, this value is also consistent with the scaling of the second moment computed directly form the Lagrangian statistic of displacements. Based on this, in the construction of the fractional model we assume $\gamma=2$, which according to Eq.~(\ref{gamma_ab}) implies $\alpha=\beta$.
This special case corresponds to the neutral fractional diffusion equation, for which $G$ in Eq.~(\ref{green_fcn}) is
\cite{mainardi_etal_2001}:
\bq
\label{eq:nfde}
G(x,t)= \frac{t^{-1}}{\pi}\frac{\sin \left[ \pi(\alpha-\zeta)/2 \right] \eta^{\alpha -1}}
{1+2\eta^{\alpha}\cos \left[ \pi(\alpha-\zeta)/2 \right ] + \eta^{2\alpha}}\, ,
\eq
for $\eta >0$
where $\eta=\delta x/t$ is the similarity variable and 
$\theta=\tan (\pi \zeta/2)/\tan (\pi \alpha /2)$.
The solution for $\eta <0$ is  obtained using the relation
$K(-\eta; \alpha, \zeta) =K(\eta; \alpha, -\zeta)$.
Figure~5 shows a comparison between the fractional diffusion solution in Eqs.~(\ref{gen_sol}) and  (\ref{eq:nfde}) with the PDF obtained  in Sec.~2 from the Lagrangian statistics of the quasigeostrophic transport problem. 

\begin{figure}[t]
\vspace*{2mm}
\begin{center}
\includegraphics[width=7.0cm]{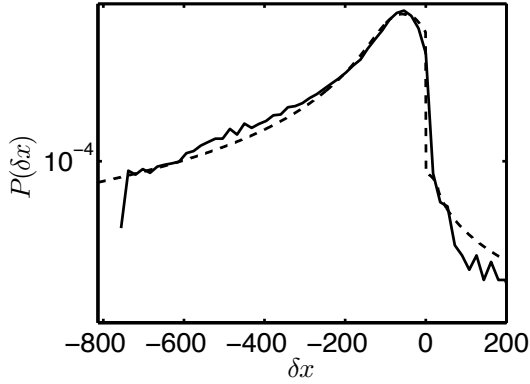}
\end{center}
\caption{Comparison between the PDF of particle displacements, $\delta x$,  in the quasigeostrophic zonal with Rossby waves(solid line),
and the PDF obtained from the solution of the fractional diffusion model in Eq.~(\ref{frac_diff}) with $\alpha=\beta=0.9$, and $\theta=1$. }
\end{figure}

\begin{figure}[t]
\vspace*{2mm}
\begin{center}
\includegraphics[width=7.0cm]{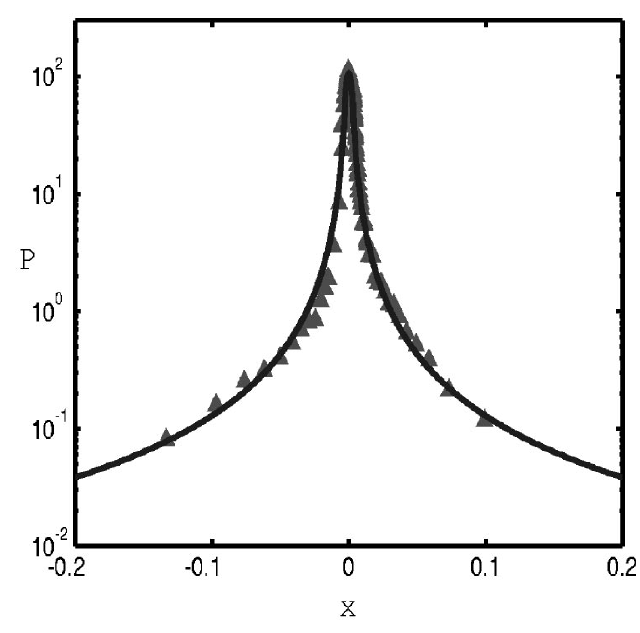}
\end{center}
\caption{Comparison between the PDF of particle displacements, $x$,  in the resistive, pressure-gradient-driven plasma turbulence model in Eqs.~(\ref{model_1})-(\ref{model_2_2}) and Fig.~4 (triangles),
and the PDF obtained from the solution of the fractional diffusion model in Eq.~(\ref{frac_diff}) with 
$\alpha=3/4$, $\beta=1/2$, $\theta=0$, and $\chi=0.09$ \cite{del_castillo_2004,del_castillo_2005}.}
\end{figure}

In the case of turbulent transport in pressure-gradient-driven plasma turbulence, the asymptotic scaling analysis of the PDFs of particle displacements according to Eqs.~(\ref{x_alpha_scaling}) and (\ref{t_beta_scaling}) and the super-diffusive scaling of the moments in
Eq.~(\ref{mo}), indicate that  $\alpha=3/4$ and $\beta=1/2$.  Figure~6 compares the solution of the fractional diffusion equation for these parameters with the PDF obtained from the direct numerical simulation shown in Fig.~4.  Details on the explicit solution of the fractional diffusion equation can be found in \cite{del_castillo_2004,del_castillo_2005}.
As discussed in Sec.~3, the Lagrangian study of transport in plasmas was based on the guiding-center equations of motion  which are an  approximation to the dynamics valid in the limit of zero Larmor radius. The role of finite Larmor radius effects on non-diffusive transport, an in particular on fractional diffusion was studied in  Ref.~\cite{kyle_2008}.

\section{Non-local transport}

In the previous sections we discussed non-diffusive transport in the context of test particle Lagrangian transport in fluids and plasmas. One of the main goals was to construct macroscopic effective transport models to describe the PDF of particle displacements in chaotic and turbulent flows. It was shown that fractional diffusion operators provide a 
framework to describe the spatio-temporal evolution of the PDFs. In particular, the  long tails of the PDFs as well as the non-Gaussian scaling of the Lagrangian statistics are well capture by fractional diffusion models. Motivated by these results, in this section we discuss the use of fractional diffusion models to describe non-diffusive transport of passive scalars, like temperature, density, pressure or the concentration of a pollutant in  flow. 

The starting point is the conservation law 
\bq
\label{continuity}
\partial_t T = -\partial_x q \, ,
\eq 
where $T$ denotes the scalar field transported and $q$ denotes the flux. For simplicity we limit attention to the transport of a single scalar in a $1$-dimensional domain. The conservation law (\ref{continuity}) has to be complemented with a prescription relating  $q$ and $T$. In the case of diffusive-transport this closure is provided by the Fourier-Fick's local prescription
\bq
\label{ff}
q=- \chi \partial_x T + V T\, ,
\eq
where $\chi$ is diffusion coefficient and $V$ is the advection velocity.
Substituting Eq.~(\ref{ff}) into  Eq.~(\ref{continuity}) leads to the advection-diffusion model in Eq.~(\ref{adv_diff}). 

Although the advection-diffusion model has been successfully applied to a wide variety of transport problems, there are cases in which this model fails to describe the dynamics. The examples discussed before showed clear evidence of this in the case of the PDF of Lagrangian particle displacements. Here we explore the role of non-diffusive transport of scalars, like temperature, for which a Lagrangian test particle perspective might not be readily available. One of the main motivations for this study is the understanding of fast propagation phenomena in magnetically confined plasmas. The basic problem can be understood without entering into the details concerning  the plasma system. The top panel in Fig. ~8 shows the basic configuration of interest, where $T_0$ is perturbed by a pulse at the edge of the domain. The problem then is to study the relaxation of the system back to the steady state. These type of perturbative transport experiments are commonly performed in magnetically confined fusion devices where a plasma is suddenly cooled at the edge. It has been observed in several  experiments that such cold pulse perturbations travel from the edge to the center of the device at speeds significantly greater than the typical diffusive time scales. Because of this, attempts to model some of these experiments using the diffusion equation  have failed. Here we discuss the use of non-local transport models as an alternative to diffusive models to describe these phenomena. 

By non-local we mean that, contrary to the Fourier-Fick's local prescription in Eq.~(\ref{ff}), the flux $q$ at a given point depends on the gradient of $T$ throughout the entire domain. The  generic mathematical structure of these nonlocal models is
\bq
\label{nl}
q(x)=- \chi \int {\cal K} (x-y) \, \partial_{y} T(y) dy \, ,
\eq
where the kernel ${\cal K}$ determines the level of non-locality. In the case when ${\cal K}=\delta(x-x')$, Eq.~(\ref{nl}) reduces to the familiar Fourier-Fick prescription in Eq.~(\ref{ff}), where for simplicity we assume $V=0$. 

Non-local transport is a problem of significant interest in plasma physics, see for example \cite{callen_kissick_1997} and references therein. 
Fux-gradient relations of the form in Eq.~(\ref{nl}) have been used in the study of parallel
electron heat transport  in magnetized plasmas (\cite{held_2001}), and in the study of transport due to
long scale-length fluctuations (\cite{yoshizawa_itoh_2003}). However, the physics behind the non-local models discussed
here is different, and it is based on the theory of non-Gaussian stochastic processes. 
Motivated by the results discussed in the previous sections, we model the non-local flux-gradient relation in Eq.~(\ref{nl}) using fractional derivative type operators of the form
\bq
\label{nl_flux}
q=-\chi(x) \left[ l _a{\cal D}_x^\alpha - r _x{\cal D}_b^\alpha \right] T \, ,
\eq
where $\chi$ can depend on $x$ and 
\bq
_a{\cal D}_x^\alpha T = \frac{1}{\Gamma(2-\alpha)} \int_a^x \frac{T'(y)-T'(a)}{\left(x-y \right)^{\alpha-1}} dy \, ,
\eq
\bq
_x{\cal D}_b^\alpha T = \frac{1}{\Gamma(2-\alpha)} \int_x^b \frac{T'(b)-T'(y)}{\left(y-x \right)^{\alpha-1}} dy \, ,
\eq
where $T'=\partial_x T$, and $l$ and $r$ are defined in Eqs.~(\ref{eq_3_29}).
Note that the operators $_a{\cal D}_x^\alpha$ and $_x{\cal D}_b^\alpha $ are not exactly the usual Riemann-Liouville fractional derivative operators introduced in Eqs.~(\ref{eq_3_30}) and (\ref{eq_3_31}). As discussed in Refs.\cite{del_castillo_2006,del_castillo_2008} this  difference has to do with the important issue that in a finite size domain $x \in (a,b)$ the  Riemann-Liouville operators must be regularized to incorporate general boundary conditions of physical interest. 


The role of non-locality and asymmetry in transport is illustrated Fig.~7 that shows the time evolution of a localized pulse initial condition in the model in Eqs.~(\ref{continuity}) and (\ref{nl_flux}) with
$\alpha=1.3$ and $\theta=0.5$. 
As the top panel shows, due to the asymmetry, $\theta \neq 0$,  the peak of the distribution shifts to the right. It can be shown that 
the peak of the profile, $x_m$, during the drift satisfies \cite{del_castillo_2006},
\bq
\label{frac_pinch_sca}
x_{m}(t)=\eta_m \,  \chi ^{1/\alpha}\,  t ^{\beta/\alpha} \, ,
\eq
where 
\bq
\label{eta_m}
\eta_m=\theta \left(\frac{\alpha+1}{2 \alpha} \right)
\alpha^{1/\alpha} \left| \tan \left( \frac{\alpha \pi}{2}\right)
\right| \, .
\eq
As expected, in the $\theta=0$ symmetric case and in the $\alpha=2$
diffusive limit, the drift vanishes.
This drift results from the existence of ``up-hill" transport which is a generic feature of non-local transport models. In the  
Fourier-Fick's prescription the flux dependence  on the local gradient is always ``down-hill", i.e., 
in the direction opposite to the local gradient. However, as the vertical lines in the top and middle panels of Fig.~7
indicate, in this case there is region of ``up-hill" transport in which the flux is in the same direction as the gradient.  
Moreover, as the bottom panel of Fig.~7 shows, in the non-local decay of the pulse, the flux-gradient relation is
 not linear like in the Fourier-Fick's diffusive case, it is in fact multivalued. 
The multivalued relation between $q$ and $-\partial_x T$ is a generic  feature of non-local transport
models with or without asymmetry. 

\begin{figure}[t]
\vspace*{2mm}
\begin{center}
\includegraphics[width=7.5cm]{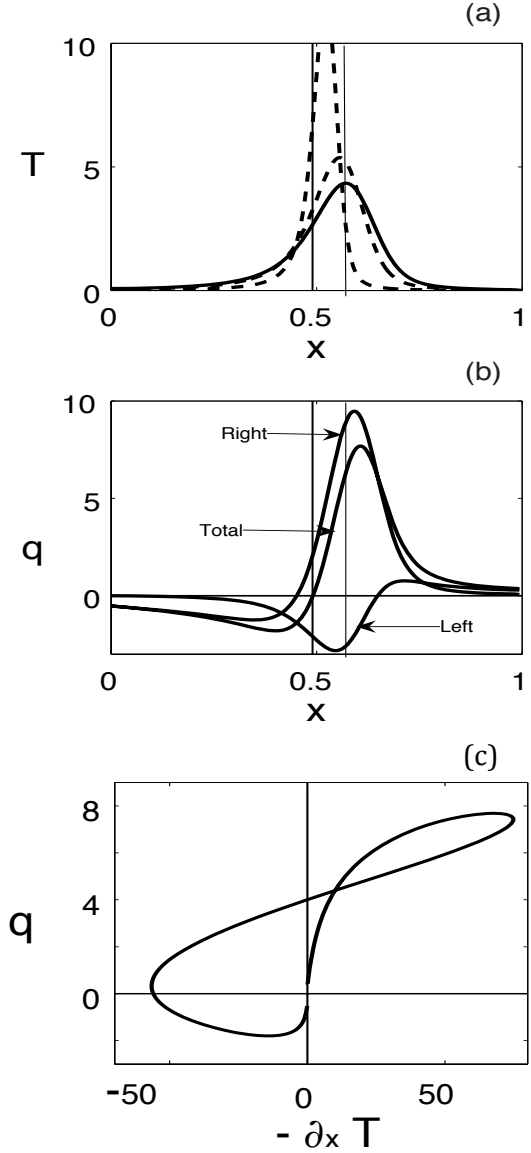}
\end{center}
\caption{
Non-local transport of a localized pulse initial condition according to the fractional diffusion model in Eqs.~(\ref{continuity}) and (\ref{nl_flux}) with 
$\alpha=1.3$, $\beta=1$, and $\theta=0.5$. The solid line in panel (a) shows the profile at the final
time, and the dashed lines the profiles at earlier times. The drift of the distribution results from the asymmetry $\theta \neq 0$ of the fractional operator. Panel (b) shows the left, $q_l$, the right, $q_r$,  and the total non-local flux, $q$, and panel (c) shows the flux-gradient relation. Contrary to the Fourier-Fick's 
 linear relation, $q = -\chi \partial_x T$,  $q$ and $-\partial_x T$ exhibit a nonlinear, multivalued relation. The top, left quadrant, $-\partial_x T <0$ and $q>0$ corresponds to up-hill transport that occurs in the region bounded by the two vertical lines in panels (a) and (b). }
\end{figure}


In the study of the propagation of pulse perturbations,
the first step  is the computation of the steady equilibrium
temperature profile, $T_0(x)$ in the presence of a source of the form
\bq
\label{eq_10}
S= S_0\, {\rm exp} \left[ -\frac{(x-\mu_s)^2}{2\sigma_s^2} \right ] \, ,
\eq
with  $\mu_s=0$, and $\sigma_s=0.075$. For each simulation, the source amplitude was selected so that $T_0(0)=1$. 
The simulations followed the spatio-temporal evolution 
of the perturbed temperature, $\delta T (x,t) = T(x,t) -
T_0(x)$, with 
initial condition 
\bq
\delta T (x,0)=- A \exp\left[ - \frac{\left( x- \mu_p\right)^2}{2 \sigma_p^2}\right]\, ,
\eq
where $A=0.3$, $\mu_p=0.75$, and $\sigma_p=0.03$.  Details on the numerical method used to integrate the fractional transport model can be found in \cite{del_castillo_2006}.
The bottom panel in Fig.~8  shows the time evolution of the normalized  tracer perturbation, $\hat {\delta T}=\delta T / |{\rm min} \left[ \delta T (x,0) \right]|$, at different locations along the $x$-domain.  We define the mean pulse propagation speed as the ratio of the normalized distance and the time delay, $V_p=1/\delta t$. The time delay is defined as the time required for the scalar field at $x=0$ to exhibit a drop of size $\delta T_c$. That is, $\delta T(0, \delta t)=\delta T_c$. For the value of the threshold we choose $\delta T_c=-0.0375$. We considered three case: an $\alpha=2$ diffusive case, and two fractional cases with $\alpha=1.75$ and $\alpha=1.25$. The main conclusion is that non-locality can lead to a considerable increase of the pulse speed. In particular, the numerical results show that for the same value of $\chi$, $V_p$ for $\alpha=1.25$ is about $10$ bigger than the diffusive speed.
This idea was used in Ref.~\cite{del_castillo_2008} to model perturbative experiments on cold temperature pulse propagation in the Join European Torus (JET) magnetically confined controlled fusion device.

To conclude we present recent results on the role of non-locality in the propagation of pulses through transport barriers. 
The local and non-local diffusivities are assumed to be of the form
\bq
\chi_d=\chi_{d 0}  
 - \zeta e^{-(x-x_0)^2/ w}  \, ,
\eq
and 
\bq
\chi_{nl}=
\frac{\chi_{nl 0}}{2}\left[
\tanh \left ( 
\frac{x-x_{c}}{L}
\right )
+ \tanh \left ( 
\frac{x_{c}}{L}
\right )
\right ] -
\eq
$$
 - \zeta e^{-(x-x_0)^2/ w}
 \, .
$$
The $\tanh$ profile in $\chi_{nl}$ is introduced to guarantee the vanishing of the non-local flux in the core region where transport is assumed to be dominated by diffusive processes. 
The transport barrier is modeled  by introducing a dip, $e^{-(x-x_0)^2/ w}$, in the diffusivity profiles. In the calculations reported here
 $\chi_{d 0}=1$, $x_0=0.5$, $\zeta=0.95$, $\chi_{nl 0}=1$, $x_c=0.1$,  $L=0.025$, and $w=0.005$. 
 In the non-local simulations, $\alpha=1.25$. 
Figure~9 shows the spatio-temporal evolution of $\delta T$.  The top panel shows the case of diffusive transport, 
$\chi_{nl0}=0$, in the absence of transport barriers. In this case, the pulse spreads throughout  the plasma domain in a slow, diffusive time scale. As expected, as shown in the middle panel, in the presence of a transport barrier the diffusive propagation of the pulse is stopped. However, in the presence of non-local transport the pulse dynamics is fundamentally different. 
As the bottom panel in Fig.9~ shows, in this case the pulse can in fact go through the transport barrier. This ``tunneling" effect is a unique novel property of non-local transport.

\begin{figure}[t]
\vspace*{2mm}
\begin{center}
\includegraphics[width=7.5cm]{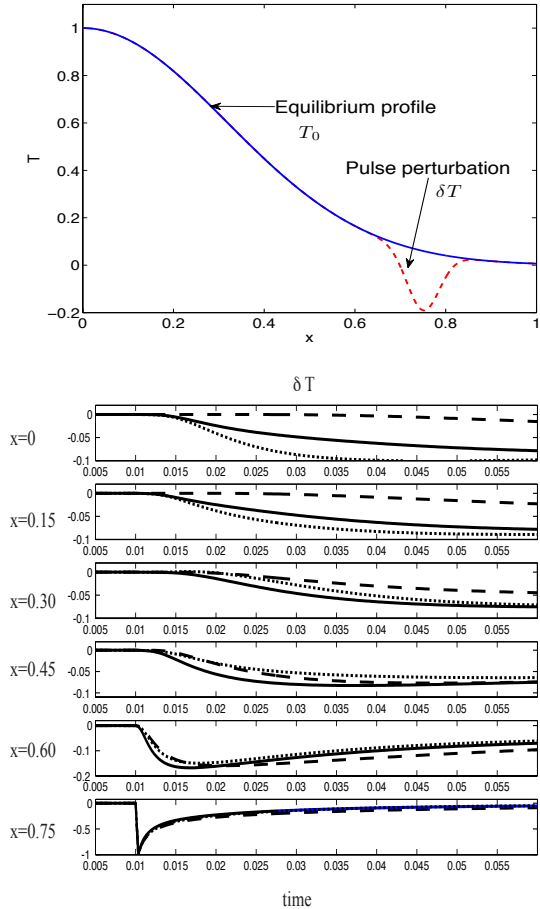}
\end{center}
\caption{Non-local fast pulse propagation. 
As shown in the top panel, 
perturbative transport simulations follow the evolution of a localized perturbation (dashed line) of an steady state passive tracer profile (solid line). The bottom panel shows the time traces of the normalized tracer perturbation, $\hat {\delta T}=\delta T / |{\rm min} \left[ \delta T (x,0) \right]|$, at different locations along the $x$ domain. 
In the local diffusive case (dashed line) the normalized propagation speed from the edge, $x=0.75$, to the center, $x=0$, of the domain is $\hat{V_p}=1$. In the fractional case with $\alpha=1.75$ (solid line), $\hat{V_p}=6.3$, and in the fractional case with 
$\alpha=1.25$ (dotted line), $\hat{V_p}=9.6$. }
\end{figure}

\begin{figure}[t]
\vspace*{2mm}
\begin{center}
\includegraphics[width=7.cm]{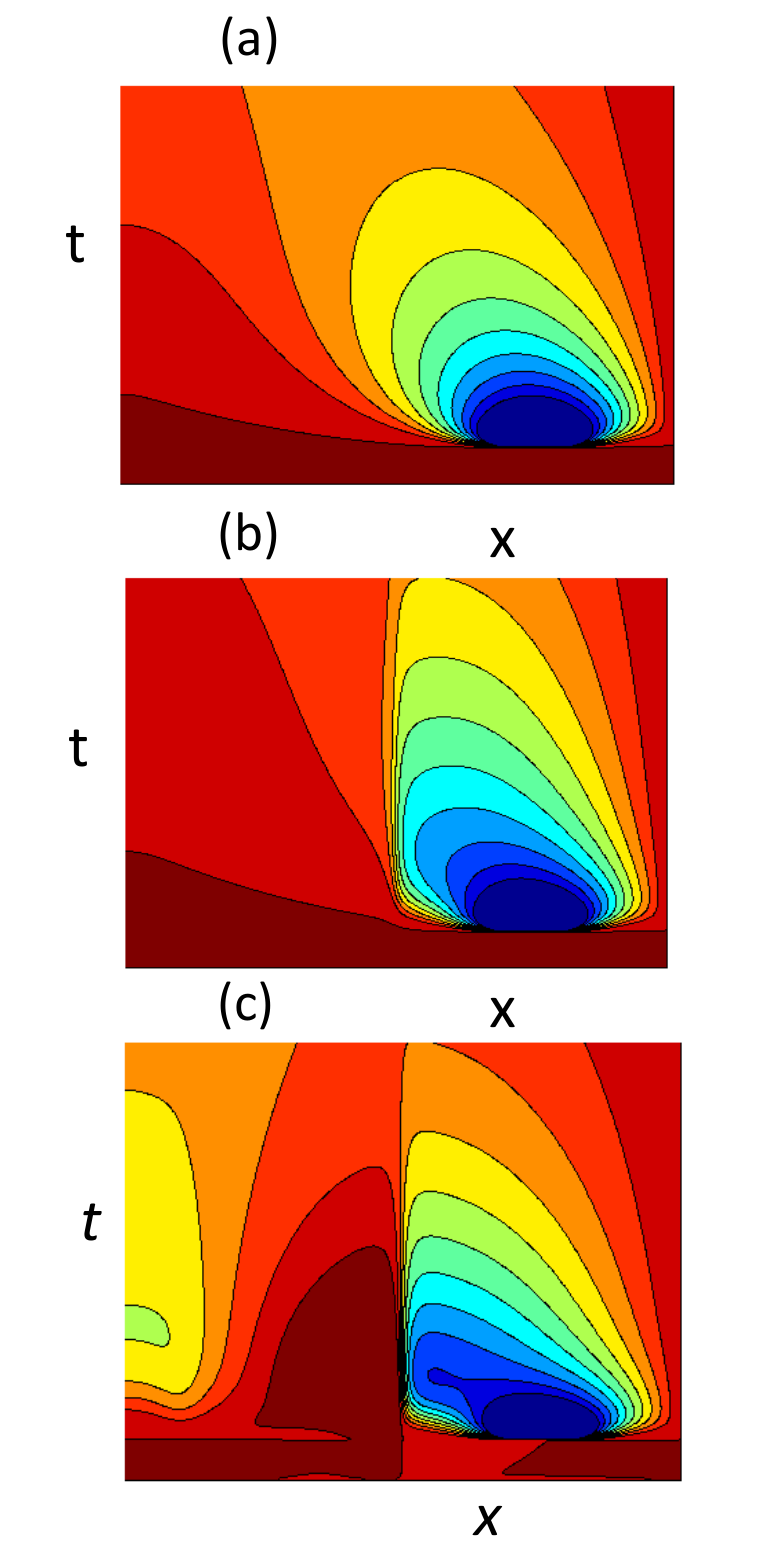}
\end{center}
\caption{Non-local ``tunneling" of perturbations across a transport barrier. The figure shows the space-time evolution of the normalized passive tracer perturbation $\hat {\delta T}=\delta T / |{\rm min} \left[ \delta T (x,0) \right]|$ with dark blue (red) denoting $\hat {\delta T}=1$ ($\hat {\delta T}=0$). The top panel corresponds to diffusive transport in the absence of  transport barriers. The middle and bottom panels correspond to diffusive and non-local transport respectively in the presence of a transport barrier. The vertical dashed line indicates the location of the transport barrier.}
\end{figure}

\conclusions

We have presented a review of recent results on non-diffusive transport in fluids and plasmas. 
The approach was based on the study of the Lagrangian statistics of large ensembles of particles. In general, the stochasticity in the Lagrangian trajectories can result from deterministic chaos or from turbulence. The examples discussed encompass both possibilities.  In the studied of transport by Rossby waves in quasigeostrophic zonal flows, the advection velocity was a smooth deterministic function but the Lagrangian trajectories exhibited Hamiltonian chaos. 
On the other hand, in the ${\bf E} \times {\bf B}$ transport plasma problem, the advection velocity was a non-deterministic random function obtained from the solution of a turbulence model. The main object of study was the probability density function (PDF) of individual particle displacements, also know as the propagator.  Both, the fluid chaotic transport problem and the plasma turbulent transport problem, exhibited strongly non-Gaussian spatio-temporal self-similar PDFs. 
In addition, the Lagrangian statistics in both cases exhibited super-diffusive scaling, $<x^2> \sim t^\gamma$ with $\gamma>1$. The modeling of these PDFs using advection-diffusion equations is out of the question because the effective diffusivity diverges, and  the propagators have non-Gaussian decaying tails. The observed non-Gaussian statistics in the examples discussed has its origin on the combination of anomalously large particle displacements, known as ``Levy flights", and the trapping effects of coherent structures like fluid vortices and ${\bf E} \times {\bf B}$ plasma eddies.  

We have shown that the PDFs of particle displacements can be modeled using fractional diffusion equations in which regular derivatives are replaced by fractional derivatives. Fractional derivatives are integro-differential operators that provide a powerful, elegant  framework to incorporate non-Gaussian and non-Markovian effects on transport models.  These operators naturally appear in the continuum limit of generalized random walk models that extend the Brownian motion by allowing non-Gaussian jump distribution functions and general waiting time distribution functions. 

Going beyond the study of non-Gaussian Lagrangian statistics, we discussed the application of fractional derivatives to model non-local transport.  The  cornerstone of the diffusive transport paradigm is the Fourier-Fick's prescription according to which the flux at a given point depends only of the gradient of the transported field at that point. On the other hand, in the case of non-local transport, the flux can depend on the gradient throughout the entire domain. Although in many cases transport problems follow the Fourier-Fick's prescription, there are important situations in which this is not the case. A clear example is the fast propagation phenomena observed in perturbative transport experiments in magnetically confined plasma fusion devices. Motivated by the successful use of fractional derivatives in the study of non-diffusive Lagrangian transport, we used these operators to construct non-local models of passive scalar transport. We presented numerical results illustrating important non-local transport phenomenology including: up-hill transport, multivalued flux-gradient relations, fast pulse propagation phenomena, and ``tunneling" of perturbations across transport barriers.  

Some of the results presented here pertain specific systems, i.e.,  Rossby waves in zonal flows and pressure-gradient-driven plasma turbulence. However, it is important to realize that the observed non-diffusive phenomenology depends on very general non-Gaussian statistical properties and not on specific details. In particular, other systems with coherent structures and/or strong spatio-temporal correlations are likely to exhibit similar non-diffusive and non-local transport dynamics. 



\begin{acknowledgements}
This work was sponsored by the Oak Ridge National Laboratory, managed
by UT-Battelle, LLC, for the U.S. Department of Energy under contract DE-AC05-00OR22725.
\end{acknowledgements}







%




%




%



\addtocounter{figure}{-1}\renewcommand{\thefigure}{\arabic{figure}a}

\end{document}